*Р. В. Майер*
*Доцент, доктор педагогических наук*

# РЕШЕНИЕ ЗАДАЧ МАТЕМАТИЧЕСКОЙ ТЕОРИИ ОБУЧЕНИЯ МЕТОДОМ КОМПЬЮТЕРНОГО МОДЕЛИРОВАНИЯ

Проанализированы модели обучения, учитывающие, что: 1) скорость увеличения знаний ученика пропорциональна разности между уровнем требований учителя и уже имеющимися знаниями; 2) при слишком высоких требованиях мотивация ученика снижается и он перестает учиться. Предложены: 1) однокомпонентная модель, исходящая из того, что учебная информация состоит из равноправных элементов; 2) двухкомпонентная модель, учитывающая, что знания усваиваются с различной прочностью, прочные знания забываются существенно медленнее непрочных; 3) двухкомпонентная модель, которая учитывает переход непрочных знаний в прочные. Приведено решение пяти прогностических и оптимизационных задач теории обучения.

Ключевые слова: дидактика, математическая теория обучения, имитационное моделирование, модель обучения, программирование, оптимизация.

*R. V. Maier*
*Associate professor, doctor of pedagogical sciences*

# SOLVE OF PROBLEMS OF MATHEMATICAL THEORY OF LEARNING WITH USING COMPUTER MODELING METHODS

Analyzed models of learning, which take into account that: 1) the rate of increase of student's knowledge is proportional to the difference between levels of teacher's requirements and prior knowledge; 2) if the requirements are too high, then student motivation decreases and he stops learning. Was proposed: 1) a one-component model, coming from the fact that the training information consists of equal elements; 2) a two–component model that takes into account that knowledge is assimilated with varying strength, "trustworthy" knowledge forgotten much slower then "weak"; 3) two-component model, which takes into account the transition of "weak" knowledge in "trustworthy" knowledge. The solution of the five predictors and optimization problems of learning theory are represented.

Математическая теория обучения (МТО), возникшая на стыке дидактики и математики, занимается исследованием процесса обучения с помощью математических методов [2]. Развитие информационных технологий создало предпосылки для использования метода имитационного моделирования с целью исследования дидактических процессов [1, 4]. Все задачи МТО можно разделить на два класса: 1) **прогностические**: зная параметры учеников, характеристики используемых методов и учебную программу (распределение учебной информации), определить их уровень знаний (или сформированности навыка) в последовательные моменты времени и в конце обучения; 2) **оптимизационные**: найти оптимальный путь обучения (применяемые методы, продолжительность занятий и т.д.), при котором уровень знаний обучающегося достигнет требуемого (или максимального) значения при заданных (или минимальных) затратах учителя и учащихся. Дальнейшее развитие МТО связано с использованием метода имитационного моделирования для решения системы задач, соответствующих тем или иным ситуациям, возникающим в процессе обучения [3, с. 52–89]. Решение каждой задачи предполагает: 1) математически строгую формулировку условия (параметры ученика, воздействие учителя, длительность занятия и т.д.); 2) выбор математической модели; 3) создание компьютерной программы, моделирующей поведение исследуемой дидактической системы; 4) проведение серии вычислительных экспериментов; 5) интерпретация и анализ результатов. Рассмотрим некоторые модели и решаемые с их помощью задачи.

**1. Однокомпонентная модель обучения.** Предположим, что сообщаемая учащимся информация (знания) является совокупностью равноправных несвязанных между собой элементов учебного материала (ЭУМ), число которых пропорционально ее количеству. Все ЭУМ одинаково легко запоминаются и с одинаковой скоростью забываются. Если уровень требований учителя превыша-ет на величину большую критического значения, то ученик перестает учиться.

Скорость увеличения знаний :

$$\frac{dZ}{dt} = \begin{cases} k\alpha Z^b(U-Z) - \gamma Z, & U \le Z + C, \\ -\gamma Z, & U > Z + C. \end{cases}$$

Здесь $\alpha$ и $\gamma$ коэффициенты научения и забывания конкретного ученика. Во время обучения ($k=1$), скорость увеличения непрочных знаний ученика пропорциональна: 1) разности между уровнем требований учителя $U$ (количеством сообщаемых знаний $z_0$) и уровнем знаний $z$ ученика; 2) количеству уже имеющихся у ученика знаний $z$ в некоторой степени $b$. Последнее позволяет учесть то, что наличие знаний способствует установлению новых ассоциативных связей и запоминанию новой информации. Когда обучение прекращается ($k=0$), количество знаний уменьшается за счет забывания. Коэффициент забывания $\gamma = 1/\tau$, где $\tau$ – время, в течение которого количество знаний уменьшается в e = 2,72 раз. Все величины измеряются в условных единицах.

**Задача 1.** Два учащихся с различными коэффициентами научения $\alpha_1 = 0,05$ и $\alpha_2 = 0,03$ изучают некоторый курс, причем уровень требований растет по закону $U = 0,0002t^2$. Надо получить графики $Z_1(t)$ и $Z_2(t)$.

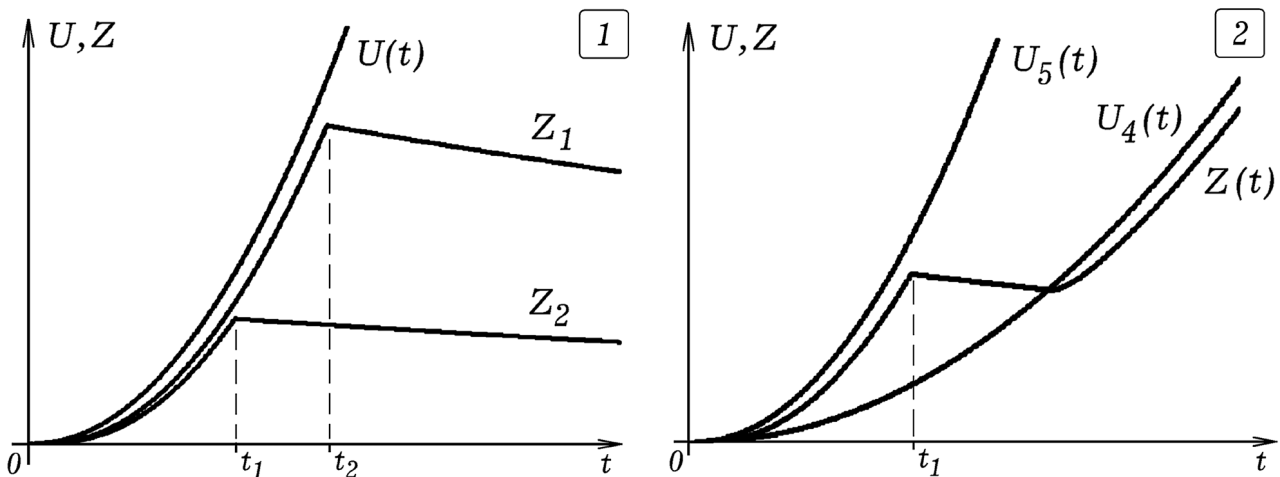

Рис. 1. Уровень требований учителя плавно увеличивается.

Имитационное моделирование дает результаты, приведенные на рис. 1.1. Сначала оба учащихся отвечают требованиям учителя (интервал [0; $t_1$] ). Уровень требований растет все быстрее и быстрее, поэтому в момент $t_1$ учащийся 2 с низким $\alpha$ отстает так сильно, что его мотивация М падает до 0, в то время как учащийся 1 продолжает соответствовать требованиям учителя. В момент $t_2$ учитель "отрывается" от обоих учеников ($U – Z > C$), предъявляя слишком высокие требования, и учащиеся перестают учиться. Учитель, заметив снижение мотивации, должен принять меры и уменьшить уровень требований $U$.

**Задача 2.** Уровни требований учителя, соответствующие оценкам 4 и 5, растут по заданным законам $U_4(t)$ и $U_5(t)$. Коэффициенты научения $\alpha$ забывания $\gamma$ ученика известны. Сначала учащийся претендует на оценку 5, но уровень требований растет слишком быстро, и поэтому он вынужден снизить уровень своих притязаний до оценки 4. Необходимо получить график $Z(t)$.

Результаты моделирования представлены на рис. 1.2. В момент $t_1$ разрыв между знаниями ученика и требованиями учителя превышает критическое значение и ученик "перестает бороться" за оценку 5. Уровень требований, соответствующий оценке 4 растет медленнее, поэтому учащийся успевает за ним.

**2. Двухкомпонентная модель обучения 1-ого типа.** С целью повышения точности результатов учтем, что прочность усвоения различных ЭУМ неодинакова, прочные знания забываются существенно медленнее нe-прочных. Рассмотрим двухкомпонентную модель ученика, при этом всю усваиваемую информацию разделим на две категории: 1) знания Зн–1, которые используются ежедневно и поэтому плохо забываются (чтение, письмо, арифметические действия, простые факты и т.д.); 2) знания Зн–2, которые применяются редко и поэтому быстро забываются (сложные идеи, принципы, факты, теории). Предлагаемая двухкомпонентная модель обучения выражается системой уравнений:

$$dZ_1 / dt = k\alpha_1 (U_1 - Z_1) Z_1^b - \gamma_1 Z_1,$$

$$dZ_2 / dt = k\alpha_2 (U_2 - Z_2) Z_2^b - \gamma_2 Z_2, \ Z = Z_1 + Z_2.$$

Здесь $U_1$ и $U_2$ – уровни требований учителя, соответствующие Зн–1 и Зн–2 , количество которых равно $Z_1$ и $Z_2$, а $Z$ – суммарные знания ученика.

**Задача 3.** Школьник в течение 11 лет учится в школе. Коэффициент усвоения информации по мере обучения увеличивается и задается матрицей $\alpha$ = (0.01, 0.015, 0.02, 0.025, 0.03, 0.035, 0.04, 0.045, 0.05, 0.055, 0.06). Уров-

ни требований учителя, соответствующие знаниям Зн–1 и Зн–2, которые необходимо усвоить в i-том классе, задаются матрицами: $U_1$ = (50, 46, 42, 36, 30, 25, 20, 15, 10, 10, 10) и $U_2$ = (4, 8, 14, 18, 24, 28, 33, 38, 46, 58, 62). Коэффициенты забывания Зн–1 и Зн–2 $\gamma_1$ = 0.002 и $\gamma_1$ = 0.01. Необходимо рассчитать суммарный уровень знаний и количество знаний Зн–1 и Зн–2 в различные моменты t.

```
Uses crt, graph;                        {PR-1: Free Pascal}
Const g1=0.002; g2=0.01; dt=0.01; Mt=2; Mz=2;
U1:array[1..11] of integer=(50,46,42,36,30,25,20,15,10,10,10);
U2:array[1..11] of integer=(4,8,14,18,24,28,33,38,46,58,62);
alfa:array[1..11] of single=(1,1.5,2,2.5,3,3.5,4,4.5,5,5.5,6);
Var t,U,Z,ZZ1,ZZ2,k: single; DV,MV,i,j: integer;
Z1,Z2: array[1..11] of single;
BEGIN DV:=Detect; InitGraph(DV,MV,'c:\bp\bgi');
Repeat t:=t+dt; U:=0; k:=1;
If (round(t) mod 12>=9)or(t>12*11-3) then k:=0;
j:=round(t) div 12 +1;
ZZ1:=0; For i:=1 to 11 do ZZ1:=ZZ1+Z1[i];
ZZ2:=0; For i:=1 to 11 do ZZ2:=ZZ2+Z2[i];
For i:=1 to 11 do begin If j=i then k:=1 else k:=0;
Z1[i]:=Z1[i]+k*alfa[i]*0.01*(U1[i]-Z1[i])*dt-g1*Z1[i]*dt;
Z2[i]:=Z2[i]+k*alfa[i]*0.01*(U2[i]-Z2[i])*dt-g2*Z2[i]*dt;
If Z1[i]<0 then Z1[i]:=0; If Z2[i]<0 then Z2[i]:=0; end;
circle(10+round(Mt*t),450-round(Mz*ZZ1),1);
circle(10+round(Mt*t),450-round(Mz*ZZ2),1);
circle(10+round(Mt*t),450-round(Mz*(ZZ2+ZZ1)),2);
circle(10+round(Mt*t),450-round(Mz*(Z1[10])),1);
circle(10+round(Mt*t),450-round(Mz*(Z2[10])),1);
until KeyPressed; CloseGraph;
END.
```

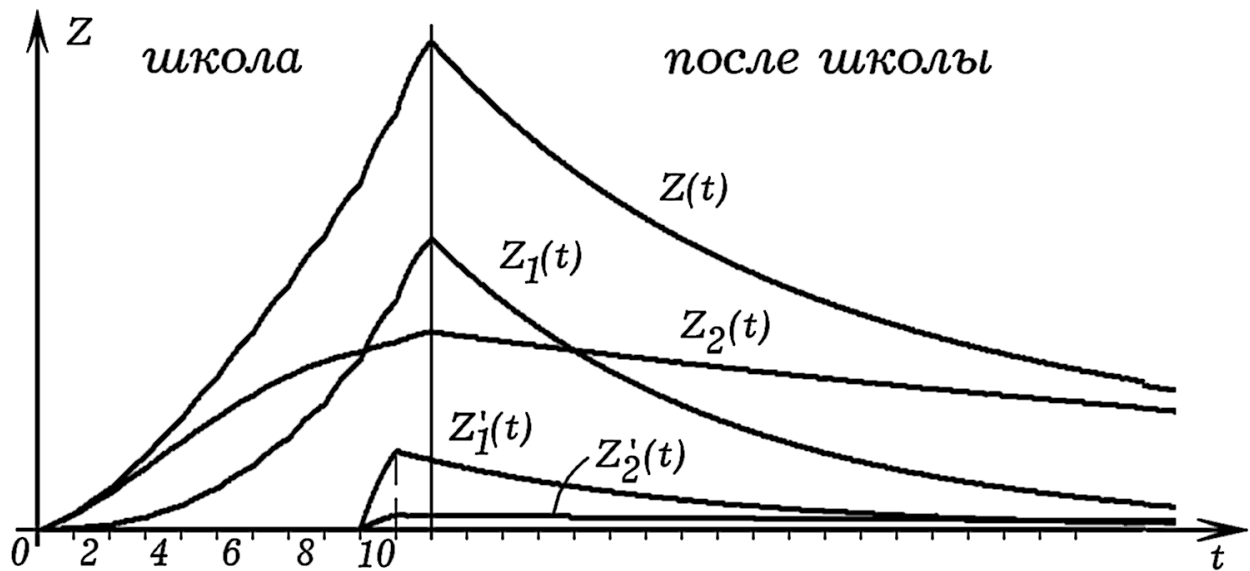

Рис. 2. Изменение количества знаний у учащегося при обучении в школе.

Используется программа PR–1, результаты – на рис. 2. На нем представлены: 1) графики $Z_1(t)$ и $Z_2(t)$ зависимостей знаний Зн–1 и Зн–2 от времени; 2) график зависимости общего количества знаний Z от времени; 3) графики $Z_1'(t)$ и $Z_2'(t)$ зависимостей знаний Зн–1 и Зн–2, приобретенных учеником в 10 классе от времени. Видно, что во время обучения в школе суммарное количество знаний, а также уровни знаний Зн–1 и Зн–2 монотонно возрастают, а после обучения убывают вследствие забывания. Знания Зн–1 забываются существенно быстрее, чем Зн–2. Условия задачи подобраны так, чтобы модель соответствовала типичной ситуации, встречающейся в педагогической практике.

3. Двухкомпонентная модель обучения 2–ого типа. Процесс усвоения и запоминания сообщаемой информации состоит в установлении ассоциативных связей между новыми и имеющимися знаниями. В результате приобретенные знания становятся более прочными и забываются значительно медленнее. Это можно учесть с помощью следующей модели:

$$dZ_1 / dt = k\alpha_1(U - Z) - k\alpha_2 Z_1 - \gamma_1 Z_1, \quad dZ_2 / dt = k\alpha_2 Z_1 - \gamma_2 Z_2, \quad Z = Z_1 + Z_2,$$

где U - уровень требований, предъявляемый учителем, и равный сообщаемым им знаниям $Z_0$, которые следует усвоить; Z – суммарное количество знаний; $Z_1$ – непрочные знания первой категории с высоким коэффициентом забывания $\gamma_1$; $Z_2$ – прочные знания второй категории с низким $\gamma_2$. Коэффициенты усвоения $\alpha_i$ характеризуют быстроту перехода знаний (i – 1)-ой категории в знания i–ой категории. Пока происходит обучение, k = 1, а когда оно прекращается k = 0. Результат обучения характеризуется не только суммарным уровнем приобретенных знаний Z, но и коэффициентом прочности Pr = $Z_2$/Z . При изучении одной темы сначала растет уровень знаний Z, затем происходит увеличение доли прочных знаний $Z_2$ и повышается прочность Pr.

Обучение будет наиболее эффективным, когда уровень требований учителя U превышает знания Z учащегося на максимально возможную величину C, при которой у учащегося еще не пропадает мотивация. Такой режим обучения называется **согласованным** или **оптимальным**. Для нахождения эффективного пути обучения, соответствующего минимальным затратам учителя или ученика, в качестве целевой функции возьмем функционал:

$$P = \int_1^2 k(U - Z)dt \approx \sum_{j=1}^n k(U_j - Z_j)\Delta t.$$

Разность U – Z характеризует интенсивность умственной деятельности (прилагаемые усилия), а величина P пропорциональна работе, совершенной учеником (или учителем). Нагрузка в течение занятия не должна превышать критическое значение $P_{max}$, чтобы избежать переутомления. Поэтому для каждого урока продолжительностью $T_i$ нужно вычислять совершенную учеником работу Pi = k(U – Z)/$\Delta t$ и сравнивать их с пороговым значением $P_{max}$.

**Задача 4.** Учащийся характеризуется параметрами $\alpha_1 = 0.01$, $\alpha_2 = 0.002$, $\gamma_1 = 0.005$ и $\gamma_2 = 0.0001$. В режиме согласованного обучения при C = 30 проводятся три занятия, начинающиеся в моменты времени 0, $t_2$=500, $t_4$=1000. Чему должна быть равна длительность занятий $T_U = t_1 = t_3$-$t_2$..., чтобы при минимальных затратах ученика уровень знаний после обучения в момент был бы не ниже , а количество знаний Z после обучения в момент t' = 1600 был бы не ниже Z' = 60, а количество знаний Z2 – не ниже 0,7Z'.

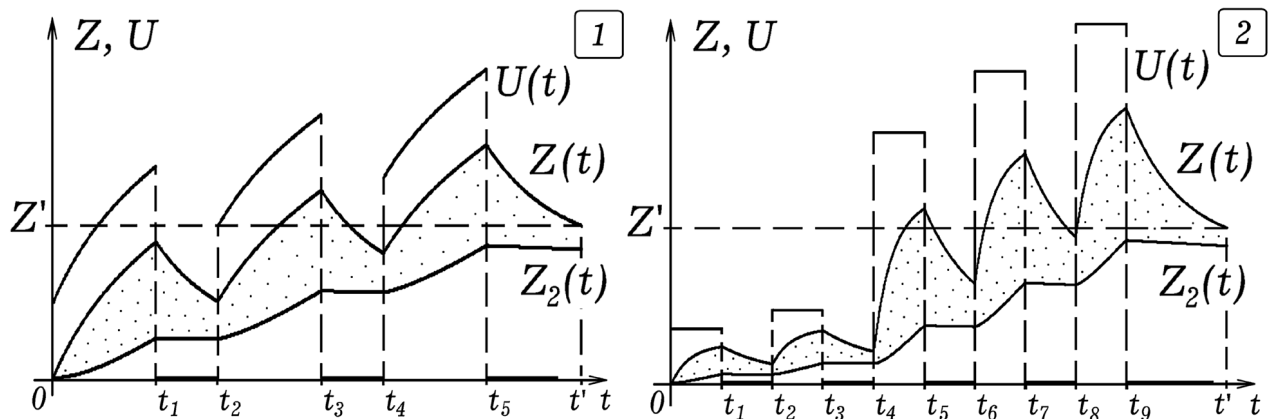

Рис. 3. Поиск оптимального пути обучения

Используемая программа содержит цикл, в котором длительность урока $T_U$ изменяется на небольшую случайную величину, и пересчитываются значения Z и $Z_2$, а также совершенная учеником работа P. Если Z > Z' и $Z_2$ > 0,7Z', а P стало меньше, то изменения $T_U$ принимаются, если нет, – отвергаются. Затем все повторяется снова. Результаты решения этой задачи представлены на рис. 3.1. Оптимальная длительность занятия составляет 312, а совершенная учеником работа – 2804.

Ученик имеет параметры $\alpha_1 = 0.01$, $\alpha_2 = 0.002$, $\gamma_1 = 0.005$ и $\gamma_2 = 0.0001$. Проводится пять занятий, начинающиеся в моменты времени 0, $t_2$=400, $t_4$=800, $t_6$=1200, $t_8$=1600, которые имеют фиксированную длительность $T_U = t_1 = t_3 - t_2... = 200$. Уровни требований, предъявляемые учителем, $U_1$, $U_2$, $U_3$, $U_4$, $U_5$ могут изменяться. Подобрать такие $U_1$, чтобы при минимальных усилиях ученика в момент t'=2200 суммарное количество его знаний Z превысило значение Z' = 90, а количество знаний второй категории $Z_2$ стало больше 0,6Z'. Работа ученика в течение урока не должна превышать $P_{max}$ = 15000.

Используемая компьютерная программа содержит процедуру, в которой рассчитываются $Z_1$, $Z_2$, Z и суммарное количество усилий учащегося P в момент t'. После этого программа случайным образом изменяет $U_i$ (i = 1, 2, ..., 5) и снова пересчитывает $Z_1$, $Z_2$, Z и P. Если новые значения удовлетворяют требованиям Z > Z' и $Z_2$ > 0,6Z', а затраченные усилия P уменьшились, то эти изменения принимаются, а в противном случае отвергаются. Результаты моделирования представлены на рис. 3.2. Программа также следит, чтобы количество усилий, затраченных на одном занятии, не превысили критического значения $P_{max}$ = 15000. Наименьшие затраты соответствуют $U_1 = 36$, $U_2 = 74$, $U_3 = 139$, $U_4 = 163$, $U_5 = 211$.

### Информация об авторе

**Майер Роберт Валерьевич** (Российская Федерация, г. Глазов) – Доцент, доктор педагогических наук, профессор кафедры физики и дидактики физики. ФГБОУ ВПО "Глазовский государственный педагогический институт им. В.Г.Короленко". Email: robert_maier@mail.ru

### Information about the author

**Maier Robert Valer'evich** (Russian Federation, Glazov) – Associate professor, doctor of pedagogical sciences, professor of the department of physics and didactic of physics, FSBEI of HPE "The Glazov Korolenko State Pedagogical Institute". Email: robert_maier@mail.ru